\documentstyle[amsfonts,eqsecnum,preprint,aps]{revtex}
\begin{document}
\draft
\title{Conformal nature of the Hawking radiation.}
\author{Massimo Materassi}
\author{materassi@pg.infn.it}
\address{Department of Physics, University of Perugia (Italy)}
\date{\today }
\maketitle

\begin{abstract}
String theory usually represents quantum black holes as systems whose
statistical mechanics reproduces Hawking's thermodynamics in a very
satisfactory way. Complicated brane theoretical models are worked out, as
quantum versions of Supergravity solutions.

These models are then assumed to be in thermal equilibrium: this is a little
cheating, because one is looking for an explanation of the seeming
thermodynamical nature of black holes, so these cannot be {\it assumed} to
be finite temperature systems!

In the model presented here, the black body spectrum arises with no
statistical hypothesis as an approximation of the unitary evolution of
microscopic black holes, which are always described by a $1+1$ conformal
field theory, characterized by some Virasoro ${\frak Vir}$ or ${\frak Vir}%
\oplus \overline{{\frak Vir}}$ algebra.

At the end, one can state that {\it the Hawking-thermodynamics of the system
is a by-product of the algebraic }${\frak Vir}\oplus \overline{{\frak Vir}}$%
{\it -symmetric nature of the event horizon}. This is {\it the central result%
} of the present work.
\end{abstract}

\bigskip

\narrowtext

\newpage  

\section{Introduction.}

When a quantum radiation in a well known initial state $\left|
in\right\rangle _{{\rm rad}}$ is sent into a black hole, then a black body
radiation is received back, so that some {\it thermalization machine} seems
to be at work, turning the quantum pure state $\left| in\right\rangle _{{\rm %
rad}}$ into the quantum mixture $\rho _{{\rm rad}}^{out}\left( T_{{\rm BH}%
}\right) $, a canonical ensemble at the Hawking temperature $T_{{\rm BH}}$.
If things work this way, the presence of black holes in our universe
immediately forces us to change drastically the basic principles of Quantum
mechanics!

More, a black hole can be attributed a finite amount of entropy, the
Beckenstein-Hawking entropy, which is shown to be an essentially geometrical
object (in classical terms), and behaves exactly as thermodynamical
entropies, tending to grow indefinitely within classical processes (when the
positive energy hypothesis is done), and decreasing when the hole
evaporates, while the total entropy of the universe grows on due to the
thermal radiation from the dying hole.

String theorists have constructed representations of the quantum states of
black holes whose evolution is assumed to be unitary. The quality check of
such stringy models consists in verifying that their quantum {\it statistical%
} mechanics fits perfectly with the values of the thermodynamical parameters
evaluated for the classical solution corresponding to them. The agreement
between these models and supergravity classical solutions is perfect (see 
\cite{Malda}, \cite{dm1}, \cite{dm2}, \cite{dm3} and many others).

This wonderful agreement, anyway, does not explain fully the mechanism of
the Hawking effect: stringy models for black holes have the right
temperature when they are supposed to be thermodynamical objects to be
treated statistically... But, why have they to be treated statistically,
even if they are describing microscopic black holes?

I think that, just as black holish thermodynamical laws and quantities
emerge outside any explicitly statistical context in the classical general
relativity, it must be possible to produce quantum models which show
thermodynamical properties without any need of postulated thermal
equilibrium: this behavior must be shared by any believable quantum model of
microscopic black holes, because it characterizes the very presence of an
event horizon.

The question addressed in this work is {\it whether it is possible to obtain
a thermal radiation from string-modeled black holes without supposing
explicitly their quantum state to be a finite temperature distribution}. The
answer seems to be that {\it it is possible}, at least for the black holes
presented here.

The paper is organized as follows.

In Section II the ''ideological'' basis of our construction is presented,
and the effective D-string model proposed for four and five dimensional
SUGRA black holes. In Section III the mathematical result is presented, and
the non-thermodynamical derivation of the Hawking effect explained. Section
IV contains a formula which relates the Hawking temperature with the
algebraic characteristics of the theory at hand, and suggests some
connection between the present results and Carlip's approach in \cite
{Carlip.3} and \cite{Carlip.4}. Section V is devoted to conclusions.

\section{The summing-averaging hiding mechanism.}

The construction of Quantum mechanics in a curved spacetime is deeply
influenced by the presence of event horizons, because such causal structures
diminish our capacity of observing the world: what we can see from outside
of the degrees of freedom and observables of the black hole, is only few
charges, the ADM mass and angular momenta.

This condition of concealment of the ''other'' quantum numbers of the black
hole is very particular: they cannot be observed, but they cannot be
''anything''! As we will see in the models presented, the structure of the
quantum Hilbert space ${\bf H}_{{\rm BH}}$ must be very well tuned to
reproduce the correct Hawking radiation. All this amount of microscopic
information sits ''behind the horizon'', but influences deeply the emitted
radiation which, even if thermal, still has a precise memory of what is
inside (in fact the $T_{{\rm BH}}$ here predicted is a function of the main
quantities defining the quantum theory).

\subsection{The horizon-centric ideology.}

Suppose to deal with a process during which the microscopic black hole emits
quanta. At the beginning of this process, the black hole quantum state is
some $\left| \psi _{{\rm BH}}\left( i\right) \right\rangle $. Then the
system emits a quantum with momentum $k$, due to some interaction $V_{{\rm %
int}}\left( k\right) $, and undergoes a transition to another quantum state $%
\left| \psi _{{\rm BH}}\left( f\right) \right\rangle $. The transition rate
of the process is evaluated in terms of the matrix element 
\begin{equation}
{\sf S}_{if}=\left\langle \psi _{{\rm BH}}\left( f\right) \right| V_{{\rm int%
}}\left( k\right) \left| \psi _{{\rm BH}}\left( i\right) \right\rangle .
\label{phdth.2}
\end{equation}

Now, the approach presented here is to consider the initial and final
quantum states of the black hole as hidden by the horizon, so the process
must be calculated as inclusive with respect to all the possible final
states with ADM quantities $\left( M^{\prime },J^{\prime },Q^{\prime
},...\right) $, and averaged over all the possible initial states with ADM
quantities $\left( M,J,Q,...\right) $; this rate reads: 
\begin{equation}
\Gamma _{f,i}^{{\rm hidden}}=%
\mathop{\displaystyle \sum }
\limits_{f,\left\langle i\right\rangle }\left| {\sf S}_{if}\right| ^2\equiv 
\frac 1{{\bf N}\left( M\right) }%
\mathop{\displaystyle \sum }
\limits_{i,f}\left| {\sf S}_{if}\right| ^2  \label{phdth.3}
\end{equation}
(here ${\bf N}\left( M\right) =\dim \ker \left( \hat M-M{\bf 1}\right) $).

The idea is than:

\begin{itemize}
\item  {\it the right quantum system modelling a given microscopic black
hole must be constructed in such a way that the quantity }$\Gamma _{f,i}^{%
{\rm hidden}}${\it \ in (\ref{phdth.3}) is a thermal distribution with the
correct values of the parameters of the corresponding classical solution}.
\end{itemize}

First of all, in order for $\Gamma _{f,i}^{{\rm hidden}}$ to be a thermal
distribution, the degeneracies must be wide enough so that averaging over
the initial states of mass equal to $M$, and summing over the final states
of mass $M^{\prime }$, is very similar to taking a thermodynamical limit in
the number of degrees of freedom \cite{Malda-N}. A black hole is thought of
as a highly excited system (see \cite{vene-dam}, \cite{Tseytlin.1} and \cite
{susskind.1}), and we need models in which ${\bf N}\left( M\right) $ is a
rapidly growing function of $M$. There are systems with this property, whose
high level degeneracy exactly matches that of black holes: in general we
will see that $1+1$ dimensional conformal field theories share this feature.

The criterion of thermal nature for $\Gamma _{f,i}^{{\rm hidden}}$ is
important because it leads to {\it a relationship between the emission
temperature of the black hole and the central charge of the CFT of the
horizon} \cite{PhDth}. This relationship seems to be important because it
appears in every model explored here (and no presentation of this ''general
coincidence'' was known to me before)\footnote{%
It is very important to say that the present results about $\Gamma _{f,i}^{%
{\rm hidden}}$ are very near to those obtained treating the systems in a 
{\it microcanonical statistical ensemble}: there appears a first degree of
approximation in which the ensemble can be considered as canonical \cite
{amarus}, while the microcanonical nature of the statement $\left| \psi _{%
{\rm BH}}\right\rangle \in \ker \left( \hat M-M{\bf 1}\right) $ is revealed
when a finer approximation is considered; at this further level, the
behavior of $\Gamma _{f,i}^{{\rm hidden}}$ resembles that of the spectrum
found in \cite{eskokeski} for the explicitly microcanonical approach.}.

If the condition of thermal nature on $\Gamma _{f,i}^{{\rm hidden}}$ holds
in the form expressed before, the correspondence point can be smoothly
reinterpreted: suppose that the system, which generated a black hole
collapsing, is a certain quantum system {\sf Q}. Typically, this will be
seen as a black hole when its ADM energy $M$ is so big that: 
\begin{equation}
r_s\left( M\right) >\ell _{{\sf Q}},  \label{phdth.4}
\end{equation}
being $\ell _{{\sf Q}}$ the physical size of {\sf Q} (see e.g. \cite
{vene-dam}). Roughly speaking, when $r_s\left( M\right) >\ell _{{\sf Q}}$ 
{\it there exists an event horizon} which (at least classically) hides the
system, allowing for the observation of ADM quantities only.

Now, the mass starts to decrease by Hawking radiation: suppose that this
process can go on till the Schwarzschild radius becomes smaller than the
physical size of {\sf Q}, as 
\begin{equation}
\ell _{{\sf Q}}>r_s\left( M\right) .  \label{phdth.5}
\end{equation}
From this point, there is {\it no event horizon}, and the external observer
can describe the system with the same precision that would be possible in
absence of any gravitational complication at all.

The picture is the following: when condition (\ref{phdth.4}) is met, many
particulars of the system are hidden by the horizon, and in this situation
any quantum process must be studied as in (\ref{phdth.3}), because ADM
quantum numbers $\left( M,J,Q\right) $ are everything that an external
observer can hope to measure.

On the contrary, when condition (\ref{phdth.5}) holds, there is no event
horizon preventing the observer to measure completely the ket $\left| \psi _{%
{\rm BH}}\right\rangle $: external observers {\it can choose} to perform an
inclusive measure, but this is simply up to them.

The correspondence principle suggests that

\begin{itemize}
\item  {\it the ''optional'' inclusive calculation performed in the case (%
\ref{phdth.5}), and its ''obligatory'' version that must be worked out in
the case (\ref{phdth.4}), have to agree at the correspondence point} 
\begin{equation}
\ell _{{\sf Q}}=r_s\left( M\right) .  \label{phdth.6}
\end{equation}
\end{itemize}

This implies that the nature of the spectrum of the (uncollapsed,
horizonless) quantum system must be such to produce a thermal spectrum when
it undergoes inclusive processes, in the suitable approximations (then, the
condition on $\Gamma _{f,i}^{{\rm hidden}}$ as expressed before); not only:
the quantities characterizing this thermodynamics, that is temperature and
entropy, must agree with the correspondent quantities for the collapsed
system in the {\it decollapsing limit} 
\[
r_s\left( M\right) \rightarrow \left( \ell _{{\sf Q}}\right) _{+}. 
\]
This conception describes the evaporation process very smoothly; anyway, it
is naive in a sense, because of {\it the key role played everywhere by the
horizon}: maybe, the fault of this model is its belief that a classical
horizon goes on existing at quantum level.

In a completely classical context, horizons are spacetime curtains that
prevent external observers from receiving any information at all from inside
the black holes; in the semiclassical context of QFT's in curved
backgrounds, horizons can stop only ordered information, because Hawking
thermal radiation can reach the future infinity.

It is expectable that in a fully quantum theory of spacetime, even ordered
information can tunnel across the horizons and reach the future infinity
(this is what results as those in \cite{eskokeski} are telling us): if that
tunnelling process were possible for ordered information, horizons would
essentially disappear in exact quantum gravity (as substantial walls can
disappear to tunnelling particles in Quantum mechanics).

\subsection{Horizon QFT for string theoretical black holes.}

A black hole is a strongly coupled system, and one could expect it is
impossible to describe it in a perturbative way; nevertheless, string theory
is able to give, at least for nearly extremal black holes, a perturbative
quantum mechanical description by turning them into {\it dual} weakly
coupled bound systems of D-branes. The curved near horizon SUGRA spacetime
is thus described by a system of quantum objects living in a flat
background, emitting and absorbing (massless) strings to which they are
weakly coupled\footnote{%
This way of working, which was found to be right in a range of cases wider
than what was originally expected, has been clarified in terms of Maldacena
conjecture and AdS/CFT correspondence.}.

In \cite{Malda} the five- and the four-dimensional SUGRA black holes are
described as dual to bound systems of D-branes, that are respectively D1-
and D5-branes in the first case, D2-, D6-branes and solitonic 5-branes in
the second;\ four or five spacelike directions are compactified, but one of
the compactification radii is much bigger than the others: 
\begin{equation}
R\gg R_i.  \label{dcb.21.bis}
\end{equation}
This fact allows for a description of such systems as {\it D-strings}.

In the low energy approximation, the starting point to study D-strings is
the {\it Born-Infeld action} \cite{Tasi} 
\begin{equation}
S_{BI}=-T_{{\rm F}}%
\displaystyle \int 
\limits_{{\Bbb V}_2^{{\rm D}}}d^2\xi e^{-\phi \left( X\right) }\sqrt{-\det
\left[ G_{{\rm mn}}\left( X\right) +B_{{\rm mn}}\left( X\right) +2\pi \alpha
^{\prime }F_{{\rm mn}}\right] },  \label{aru.92}
\end{equation}
that reduces simply to the Nambu-Goto-like functional 
\begin{equation}
S_{BI}=-T_{{\rm F}}%
\displaystyle \int 
\limits_{{\Bbb V}_2^{{\rm D}}}d^2\xi e^{-\phi _0}\sqrt{-\det \left[ G_{{\rm %
mn}}\left( X\right) \right] }  \label{aru.96}
\end{equation}
when $B_{{\rm mn}}=0$ and $F_{{\rm mn}}=0$.

Let us study the case of emission of a{\it \ gravitational wave}, with
string frame metric 
\begin{equation}
G_{\mu \nu }=e^{\frac 12\phi _0}g_{\mu \nu }=e^{\frac 12\phi _0}\left( \eta
_{\mu \nu }+2\kappa h_{\mu \nu }\right) ,  \label{aru.97}
\end{equation}
where the components of $h$ are very small, and $T_{{\rm mn}}$ is the
pull-back of a tensor $T_{\mu \nu }$ along the worldsheet of the D-string as 
$T_{{\rm mn}}=T_{\mu \nu }\partial _{{\rm m}}X^\mu \partial _{{\rm n}}X^\nu $%
.

The D-string action is then expanded in the gravitational field as follows: 
\[
\begin{array}{l}
S_{BI}=-%
{\displaystyle {T_{{\rm F}} \over \sqrt{g_{{\rm closed}}}}}
\displaystyle \int 
\limits_{{\Bbb V}_2^{{\rm D}}}d^2\xi \sqrt{-\det \eta _{{\rm mn}}\left(
X\right) }+ \\ 
\\ 
+%
{\displaystyle {T_{{\rm F}} \over \sqrt{g_{{\rm closed}}}}}
\displaystyle \int 
\limits_{{\Bbb V}_2^{{\rm D}}}d^2\xi 
{\displaystyle {\kappa \over \sqrt{-\det \eta _{{\rm mn}}\left( X\right) }}}
\left[ \eta _{\sigma \sigma }\left( X\right) h_{\tau \tau }\left( X\right)
-2\eta _{\tau \sigma }\left( X\right) h_{\tau \sigma }\left( X\right) +\eta
_{\tau \tau }\left( X\right) h_{\sigma \sigma }\left( X\right) \right] +...
\end{array}
\]

Let us now suppose $X$ to be a small fluctuation around some {\it classical
solution} $X_0${\it \ of the equations of motion }$\ddot X_0-X_0^{\prime
\prime }=0$: this $X_0$ is chosen so that 
\begin{equation}
\begin{array}{cccc}
X_0^0=\tau , & X_0^1=\sigma , & \vec X_0=\vec X_0\left( \sigma ,\tau \right)
, & \sqrt{-\det \eta _{{\rm mn}}\left( X_0\right) }=1.
\end{array}
\label{aru.102.bis}
\end{equation}
The ''exact'' field $X$ differs from $X_0$ only for {\it small fluctuations}%
\footnote{%
In this position the power $p$ of $X-X_0={\Bbb O}\left( h^p\right) $ is
chosen in the spirit of neglecting $o\left( h\right) $ everywhere.} $X-X_0=%
{\Bbb O}\left( h^{\frac 12}\right) $, the consistent approximation reads: 
\begin{equation}
\begin{array}{l}
S_{BI}=-%
{\displaystyle {T_{{\rm F}} \over \sqrt{g_{{\rm closed}}}}}
\displaystyle \int 
\limits_{{\Bbb V}_2^{{\rm D}}}d^2\xi \sqrt{-\det \eta _{{\rm mn}}\left(
X\right) }+ \\ 
\\ 
+%
{\displaystyle {T_{{\rm F}}\kappa  \over \sqrt{g_{{\rm closed}}}}}
\displaystyle \int 
\limits_{{\Bbb V}_2^{{\rm D}}}d^2\xi \left[ \eta _{\sigma \sigma }\left(
X_0\right) h_{\tau \tau }\left( X\right) -2\eta _{\tau \sigma }\left(
X_0\right) h_{\tau \sigma }\left( X\right) +\eta _{\tau \tau }\left(
X_0\right) h_{\sigma \sigma }\left( X\right) \right] +...
\end{array}
\label{aru.103.bis.bis}
\end{equation}
This action is formally the same as that of a free string perturbed by some
interaction piece 
\begin{equation}
S_{{\rm int}}\left[ X\right] =-%
{\displaystyle {T_{{\rm F}}\kappa  \over \sqrt{g_{{\rm closed}}}}}
\displaystyle \int 
\limits_{{\Bbb V}_2^{{\rm D}}}d^2\xi h_{\mu \nu }\left( X\right) \left( \dot 
X^\mu \dot X^\nu -X^{\prime \mu }X^{\prime \nu }\right) ,  \label{aru.106}
\end{equation}
which does coincide formally with the {\it graviton emission vertex for the
elementary closed string theory} \cite{GSW}. The Nambu-Goto string is
quantized {\it \`a la Polyakov}, by replacing its radical action with a
quadratic one that is completely identical to the F-string action: 
\begin{equation}
\begin{array}{cc}
S_{{\rm free}}\left[ X\right] =%
{\displaystyle {T_{{\rm D}} \over 2}}
\displaystyle \int 
\limits_{{\Bbb V}_2^{{\rm D}}}d^2\xi \left( \dot X^\mu \dot X_\mu -X^{\prime
\mu }X_\mu ^{\prime }\right) , & 
{\displaystyle {T_{{\rm F}} \over \sqrt{g_{{\rm closed}}}}}
=T_{{\rm D}}.
\end{array}
\label{aru.108.II}
\end{equation}
The $D$ spacetime components of the field $X^\mu $ are redundant, and this
is fixed by assuming 
\[
\begin{array}{cc}
X^0=\tau , & X^1=\sigma
\end{array}
\]
(static gauge), leaving $D-2$ independent variables only.

\section{Quasi thermal emissions.}

The idea that the Hawking radiation can be obtained as a purely microscopic,
non thermodynamical, effect by using $1+1$ CFT's occurred to us\footnote{%
This work is partially the synthesis of my PhD thesis \cite{PhDth}, whose
advisor was Professor Giorgio Immirzi, of the State University of Perugia.}
studying a very interesting phenomenon (discovered by D. Amati and J.G.
Russo in 1999, see \cite{amarus}) in which {\it a unitarily evolving quantum
system emits thermal radiation}, when inclusive rates are evaluated. This
system is {\it the fundamental bosonic string}: such rates\ will be referred
to as {\it quasi-thermal elementary string emission} (QUATESE for short).

\subsection{Quasi-thermal emission from elementary strings.\label
{quatese-originale}}

The action of an elementary bosonic string is: 
\begin{equation}
S=\frac{T_{{\rm F}}}2%
\displaystyle \int 
\limits_{{\Bbb V}_2}d^2\sigma \eta ^{{\rm ab}}\partial _{{\rm a}}X^\mu
\partial _{{\rm b}}X_\mu .  \label{intro.5}
\end{equation}
In open bosonic string theory the mass is quantized and each single state is
characterized by the {\it partition} $\left\{ N_n\right\} $: 
\begin{equation}
\begin{array}{cc}
M^2=%
{\displaystyle {1 \over \alpha ^{\prime }}}
\left( N-1\right) , & 
\mathop{\displaystyle \sum }
\limits_nnN_n=N.
\end{array}
\label{aru.1}
\end{equation}
It is seen that: 
\begin{equation}
\begin{array}{ccc}
N\gg 1 & \Rightarrow & \dim \ker \left( \hat N-N{\bf 1}\right) =p_N^{\left(
d\right) }\simeq 
{\displaystyle {K \over N^{\frac 14\left( d+3\right) }}}
e^{\pi \sqrt{\frac{2Nd}3}}.
\end{array}
\label{aru.3.b.ter}
\end{equation}
If the spacetime dimension is $D$, the number $d$ of {\it free independent
string fields} is $d=D-2$.

Let an open string decay from the level $N$ emitting a photon of momentum $k$
and energy $\omega $, down to the level $N^{\prime }\neq N$. The transition
takes place due to the insertion of a photon vertex $V\left( k,\xi \right)
=\xi _\mu \dot X^\mu \left( 0\right) e^{ik\cdot X\left( 0\right) }$, being $%
\xi $ the photon polarization. The decay rate is: 
\begin{equation}
\begin{array}{cc}
d\Gamma _{i,k,f}=\left| \left\langle f\right| V\left( k,\xi \right) \left|
i\right\rangle \right| ^2V\left( S^{D-2}\right) \omega ^{D-3}d\omega , & 
V\left( S^{D-2}\right) =%
{\displaystyle {2\pi ^{\frac{D-1}2} \over \Gamma \left( \frac{D-1}2\right) }}
.
\end{array}
\label{aru.7}
\end{equation}
If the only observed outputs are the photon $\left| \xi ,k\right\rangle $
and the masses of the final and of the initial states, the whole decay
probability is the sum of the $\left| \left\langle f\right| V\left( k,\xi
\right) \left| i\right\rangle \right| ^2$'s over all the final states, and
the average over the initial states of the assigned levels. This is an {\it %
inclusive decay rate}: 
\begin{equation}
d\Gamma _i\left( \omega \right) =\frac 1{p_N^{\left( d\right) }}%
\mathop{\displaystyle \sum }
\limits_{i,f}\left| \left\langle f\right| V\left( k,\xi \right) \left|
i\right\rangle \right| ^2V\left( S^{D-2}\right) \omega ^{D-3}d\omega .
\label{aru.10}
\end{equation}

The evaluation of this $d\Gamma _i\left( \omega \right) $ is performed under
the conditions 
\begin{equation}
\begin{array}{ccc}
N,N^{\prime }\gg 1, & N-N^{\prime }\ll N,N^{\prime }, & N-N^{\prime }\ll 
\sqrt{N},\sqrt{N^{\prime }},
\end{array}
\label{limite.canonico}
\end{equation}
that will be referred to as {\it canonical limit}; the very interesting
result is that $d\Gamma _i\left( \omega \right) $ has a black body form: 
\begin{equation}
\begin{array}{cc}
d\Gamma _i\left( \omega \right) =2\left( \xi ^{*}\cdot \xi \right) \alpha
^{\prime }MV\left( S^{D-2}\right) \omega ^{D-2} 
{\displaystyle {\exp \left( -\frac \omega {T_H}\right)  \over 1-\exp \left( -\frac \omega {T_H}\right) }}
d\omega , & T_H=%
{\displaystyle {1 \over 2\pi \sqrt{\alpha ^{\prime }}}}
\sqrt{%
{\displaystyle {6 \over D-2}}
}.
\end{array}
\label{aru.48.8}
\end{equation}

Amati and Russo's work explains that the emission rate $d\Gamma _i\left(
\omega \right) $ becomes a thermal rate as soon as the summing-averaging
operation $\sum_{f,\left\langle i\right\rangle }$ is performed, in the
determinant condition (\ref{limite.canonico}): a more detailed study of the
physical sense of (\ref{limite.canonico}) is going to be presented as soon
as possible in \cite{PhDth}.

The origin of this thermal nature is the wide degeneracy of the string
levels, expressed in (\ref{aru.3.b.ter}), which turns out to be an effect of
the $1+1$ conformal symmetry dominating the string physics. That the
quasi-thermal\footnote{%
I say ''quasi-thermal'' because $d\Gamma _i\left( \omega \right) $ is a
Planckian distribution only in the canonical limit (\ref{limite.canonico}).}
behavior of the elementary string emission is a by-product of the Virasoro
symmetry is written explicitly in the form of the parameter $T_H$ (check
equation (\ref{aru.48.8})): $D-2$ is {\it the number of free independent
fields}, and it can be directly equated to {\it the central charge of the
algebra }${\frak Vir}${\it \ in the CFT with fixed lightcone gauge} \cite
{GSW} 
\begin{equation}
\begin{array}{cc}
T_H=%
{\displaystyle {2 \over \ell}}
\sqrt{%
{\displaystyle {6 \over {\sf C}_{{\rm free}}}}
}, & \ell =4\pi \sqrt{\alpha ^{\prime }}.
\end{array}
\label{aru.54.quater}
\end{equation}

The rate for the emission of a soft massless particle with momentum $k$ from
a {\it closed string}, decaying from an initial state in $\ker \left( \hat N%
_R-N_R{\bf 1}\right) \otimes \ker \left( \hat N_L-N_L{\bf 1}\right) $ to a
final one in $\ker \left( \hat N_R-N_R^{\prime }{\bf 1}\right) \otimes \ker
\left( \hat N_L-N_L^{\prime }{\bf 1}\right) $, is calculated immediately
from what has been done for the open string, due to the worldsheet-chiral
factorization.

The mass shell relationship reads: 
\begin{equation}
\begin{array}{cc}
\alpha ^{\prime }M^2=4\left( N_R-1\right) +\alpha ^{\prime }Q_{+}^2=4\left(
N_L-1\right) +\alpha ^{\prime }Q_{-}^2, & N_R-N_L=m_0w_0,
\end{array}
\label{aru.72}
\end{equation}
where $m_0$ and $w_0$ are the Kaluza-Klein and winding numbers of the string
along the compactified dimension of radius $R$, and: 
\begin{equation}
Q_{\pm }=\frac{m_0}R\pm \frac{w_0R}{\alpha ^{\prime }}.  \label{aru.71}
\end{equation}
Due to the chiral factorization of states and vertices the scattering matrix
reads: 
\begin{equation}
{\sf S}_{if}=\left\langle R_f\right| V_R\left( 
{\textstyle {k \over 2}}
,\xi ^R\right) \left| R_i\right\rangle \left\langle L_f\right| V_L\left( 
{\textstyle {k \over 2}}
,\xi ^L\right) \left| L_i\right\rangle ={\sf S}_{if}^R{\sf S}_{if}^L,
\label{aru.76}
\end{equation}
so that the summed-averaged emission rate is 
\begin{equation}
\mathop{\displaystyle \sum }
\limits_{f,\left\langle i\right\rangle }d\Gamma _{i,k,f}=%
\mathop{\displaystyle \sum }
\limits_{f^{\prime },\left\langle i^{\prime }\right\rangle }\left| {\sf S}%
_{i^{\prime }f^{\prime }}^R\right| ^2%
\mathop{\displaystyle \sum }
\limits_{f^{\prime \prime },\left\langle i^{\prime \prime }\right\rangle
}\left| {\sf S}_{i^{\prime \prime }f^{\prime \prime }}^L\right| ^2V\left(
S^{D-2}\right) \omega ^{D-3}d\omega .  \label{aru.77.bis}
\end{equation}

Step by step, what has been done for the open string can be repeated for the
two chiral halves, getting: 
\begin{equation}
\mathop{\displaystyle \sum }
\limits_{f,\left\langle i\right\rangle }\left| {\sf S}_{if}^{R,L}\right|
^2\simeq K%
{\displaystyle {\omega \over 2}}
\left( \xi _{R,L}^{*}\cdot \xi _{R,L}\right) M\alpha ^{\prime } 
{\displaystyle {\exp \left( -\frac \omega {2T_{R,L}}\right)  \over 1-\exp \left( -\frac \omega {2T_{R,L}}\right) }}
.  \label{aru.79.bis}
\end{equation}
These are {\it black body spectra}: the left- and the right-moving
temperatures are: 
\begin{equation}
\begin{array}{cc}
T_L=%
{\displaystyle {\sqrt{M^2-Q_{-}^2} \over a\sqrt{\alpha ^{\prime }}M}}
, & T_R=%
{\displaystyle {\sqrt{M^2-Q_{+}^2} \over a\sqrt{\alpha ^{\prime }}M}}
\end{array}
\label{aru.81}
\end{equation}
and can be expressed in terms of the central charges as: 
\begin{equation}
\begin{array}{cc}
T_L=%
{\displaystyle {1 \over \pi \alpha ^{\prime }M}}
\sqrt{%
{\displaystyle {6\left( N_L-1\right)  \over {\sf C}_{{\rm free}}}}
}, & T_R=%
{\displaystyle {1 \over \pi \alpha ^{\prime }M}}
\sqrt{%
{\displaystyle {6\left( N_R-1\right)  \over {\sf \tilde C}_{{\rm free}}}}
}.
\end{array}
\label{aru.88.bis}
\end{equation}
By approximating $M^{-1}$ as the $M_0^{-1}$ of the unexcited string at rest,
with $M_0=L_{{\rm F}}T_{{\rm F}}$ and $T_{{\rm F}}=\frac 1{2\pi \alpha
^{\prime }}$, one can write: 
\begin{equation}
\begin{array}{cc}
T_L=%
{\displaystyle {2 \over L_{{\rm F}}}}
\sqrt{%
{\displaystyle {6\left( N_L-1\right)  \over {\sf C}_{{\rm free}}}}
}, & T_R=%
{\displaystyle {2 \over L_{{\rm F}}}}
\sqrt{%
{\displaystyle {6\left( N_R-1\right)  \over {\sf \tilde C}_{{\rm free}}}}
}.
\end{array}
\label{aru.88.ter}
\end{equation}
The closed string decay rate reads: 
\begin{equation}
\begin{array}{l}
d\Gamma = 
{\displaystyle {K^2\left( \xi _R^{*}\cdot \xi _R\right) \left( \xi _L^{*}\cdot \xi _L\right) M^2\alpha ^{\prime 2} \over 4}}
\\ 
\\ 
{\displaystyle {\exp \left( -\frac \omega {2T_R}\right)  \over \left[ 1-\exp \left( -\frac \omega {2T_R}\right) \right] }}
{\displaystyle {\exp \left( -\frac \omega {2T_L}\right)  \over \left[ 1-\exp \left( -\frac \omega {2T_L}\right) \right] }}
V\left( S^{D-2}\right) \omega ^{D-1}d\omega .
\end{array}
\label{aru.83}
\end{equation}

This decay rate can be rewritten as the product of a black body spectrum
with temperature: 
\begin{equation}
T^{-1}=\frac 12\left( T_R^{-1}+T_L^{-1}\right) ,  \label{aru.84}
\end{equation}
times the suitable {\it grey body factor}: 
\begin{equation}
\sigma \left( \omega ,T\right) = 
{\displaystyle {1-\exp \left( -\frac \omega T\right)  \over \left[ 1-\exp \left( -\frac \omega {2T_R}\right) \right] \left[ 1-\exp \left( -\frac \omega {2T_L}\right) \right] }}
\omega .  \label{aru.85}
\end{equation}

\subsection{QUATESE for D-brane modeled black holes.\label
{quatese-D-stringhe}}

The low energy physics of a D-string is governed by actions which are
formally equal to those of the F-string. The naive conjecture is that
everything could work the same way for the D- as for the F-string, with the
replacement: 
\begin{equation}
T_{{\rm F}}\rightarrow T_{{\rm D}}.  \label{aru.108.bis}
\end{equation}
If this system decays from a very excited level to some slightly less
energetic level, the emission will be quasi thermal again, when summed over
the final polarizations and averaged over the initial ones.

Unfortunately things are not so simple: the elementary string mass levels in
(\ref{aru.72}) are not those of a D-string. The spectrum of non BPS states
of a D-superstring reads \cite{dm1}: 
\begin{equation}
\left\{ 
\begin{array}{l}
M_{{\rm D}}\simeq LT_{{\rm D}}+%
{\displaystyle {4\pi  \over L}}
\left( N_R-\delta _R\right) -%
{\displaystyle {2\pi  \over L}}
m_0w_0, \\ 
\\ 
M_{{\rm D}}\simeq LT_{{\rm D}}+%
{\displaystyle {4\pi  \over L}}
\left( N_L-\delta _L\right) +%
{\displaystyle {2\pi  \over L}}
m_0w_0,
\end{array}
\right.  \label{aru.112}
\end{equation}
(being $L$ the length of the D-string\footnote{%
The approximated equality symbol $\simeq $ is due to the fact that this
would be exact only if the massless open strings travelling along the
D-string did not interact.} and $\delta _{R,L}$ corrective constant zero
point terms due to the supersymmetry), while the elementary string spectrum
gives mass levels of the form: 
\begin{equation}
\left\{ 
\begin{array}{l}
M_{{\rm F}}=\sqrt{8\pi T_{{\rm F}}\left( N_R-\delta _R\right) +%
{\displaystyle {4\pi ^2w_0^2m_0^2 \over L^2}}
+4\pi T_{{\rm F}}m_0w_0+L^2T_{{\rm F}}^2}, \\ 
\\ 
M_{{\rm F}}=\sqrt{8\pi T_{{\rm F}}\left( N_L-\delta _L\right) +%
{\displaystyle {4\pi ^2w_0^2m_0^2 \over L^2}}
-4\pi T_{{\rm F}}m_0w_0+L^2T_{{\rm F}}^2}.
\end{array}
\right.  \label{aru.113}
\end{equation}

In the {\it long string hypothesis }$L\sqrt{T_{{\rm F}}}\gg 1$ the mass in (%
\ref{aru.113}) can be expanded as: 
\begin{equation}
\left\{ 
\begin{array}{l}
M_{{\rm F}}=T_{{\rm F}}L+%
{\displaystyle {2\pi  \over L}}
m_0w_0+%
{\displaystyle {4\pi  \over L}}
\left( N_R-\delta _R\right) +{\Bbb O}\left( 
{\displaystyle {1 \over T_{{\rm F}}L^3}}
\right) , \\ 
\\ 
M_{{\rm F}}=T_{{\rm F}}L-%
{\displaystyle {2\pi  \over L}}
m_0w_0+%
{\displaystyle {4\pi  \over L}}
\left( N_L-\delta _L\right) +{\Bbb O}\left( 
{\displaystyle {1 \over T_{{\rm F}}L^3}}
\right) ,
\end{array}
\right.  \label{aru.115}
\end{equation}
so in this regime there exists a perfect duality between the elementary
string mass and the D-string one (this is the S-duality presented in \cite
{Schwarz.1}).

More rigorously, let us consider what changes in Amati and Russo's argument
if the F-string is replaced by such a D-string. The quantum numbers $N_{R,L}$
are: 
\begin{equation}
\left\{ 
\begin{array}{l}
N_R=%
{\displaystyle {L \over 4\pi }}
M_{{\rm D}}-%
{\displaystyle {L^2 \over 4\pi }}
T_{{\rm D}}-%
{\displaystyle {w_0m_0 \over 2}}
+\delta _R, \\ 
\\ 
N_L=%
{\displaystyle {L \over 4\pi }}
M_{{\rm D}}-%
{\displaystyle {L^2 \over 4\pi }}
T_{{\rm D}}+%
{\displaystyle {w_0m_0 \over 2}}
+\delta _L.
\end{array}
\right.  \label{aru.118.b}
\end{equation}
The expressions of $N_{R,L}^{\prime }$ in terms of $M_{{\rm D}}^{\prime }$
are formally equal ($w_0$ and $m_0$ do not change). With some kinematics, $%
M_{{\rm D}}^{\prime }$ is related to $M_{{\rm D}}$ (the D-string being
initially at rest), and for small $\omega \ $one has: 
\begin{equation}
\begin{array}{cc}
N_R^{\prime }=N_R-%
{\displaystyle {L \over 4\pi }}
\omega , & N_L^{\prime }=N_L-%
{\displaystyle {L \over 4\pi }}
\omega .
\end{array}
\label{aru.118.l}
\end{equation}

The inclusive process transition rate reads: 
\[
\mathop{\displaystyle \sum }
\limits_{f,\left\langle i\right\rangle }d\Gamma _{i,\omega ,f}=%
\mathop{\displaystyle \sum }
\limits_{f^{\prime },\left\langle i^{\prime }\right\rangle }\left| {\sf S}%
_{i^{\prime }f^{\prime }}^R\right| ^2%
\mathop{\displaystyle \sum }
\limits_{f^{\prime \prime },\left\langle i^{\prime \prime }\right\rangle
}\left| {\sf S}_{i^{\prime \prime }f^{\prime \prime }}^L\right| ^2V\left(
S^{D-2}\right) \omega ^{D-3}d\omega , 
\]
due to the chiral factorization: the rest is pure mathematics, the net
result is again: 
\begin{equation}
\mathop{\displaystyle \sum }
\limits_{f^{\prime },\left\langle i^{\prime }\right\rangle }\left| {\sf S}%
_{i^{\prime }f^{\prime }}^{R,L}\right| ^2=K%
{\displaystyle {L\omega  \over 4\pi }}
\left( \xi _{R,L}^{*}\cdot \xi _{R,L}\right) 
{\displaystyle {\exp \left( -\frac \omega {2T_{R,L}}\right)  \over 1-\exp \left( -\frac \omega {2T_{R,L}}\right) }}
,  \label{aru.118.m}
\end{equation}
a quasi-thermal emission with the temperatures: 
\begin{equation}
\begin{array}{cc}
T_R^{\left( {\rm D}\right) }=%
{\displaystyle {2 \over L}}
\sqrt{%
{\displaystyle {6N_R \over D-2}}
}, & T_L^{\left( {\rm D}\right) }=%
{\displaystyle {2 \over L}}
\sqrt{%
{\displaystyle {6N_L \over D-2}}
}.
\end{array}
\label{aru.118.n}
\end{equation}
By involving the central charges ${\sf C}_{{\rm free}}^{\left( {\rm D}%
\right) }={\sf \tilde C}_{{\rm free}}^{\left( {\rm D}\right) }=D-2$, this
relationship reads: 
\begin{equation}
\begin{array}{cc}
T_R^{\left( {\rm D}\right) }=%
{\displaystyle {2 \over L}}
\sqrt{%
{\displaystyle {6N_R \over {\sf C}_{{\rm free}}^{\left( {\rm D}\right) }}}
}, & T_L^{\left( {\rm D}\right) }=%
{\displaystyle {2 \over L}}
\sqrt{%
{\displaystyle {6N_L \over {\sf \tilde C}_{{\rm free}}^{\left( {\rm D}\right) }}}
}.
\end{array}
\label{temps-D-stringa}
\end{equation}

The grey body spectrum can be re-constructed as 
\begin{equation}
\frac{d\Gamma }{V\left( S^{D-2}\right) d\omega }=K^2\left( \xi _R^{*}\cdot
\xi _R\right) \left( \xi _L^{*}\cdot \xi _L\right) 
{\displaystyle {L^2 \over 16\pi ^2}}
\omega ^{D-2}\sigma \left( \omega ,T_{{\rm BH}}\right) 
{\displaystyle {\exp \left( -\frac \omega {T_{{\rm BH}}}\right)  \over 1-\exp \left( -\frac \omega {T_{{\rm BH}}}\right) }}
\label{aru.118.o}
\end{equation}
with 
\begin{equation}
\begin{array}{cc}
\sigma \left( \omega ,T\right) = 
{\displaystyle {\left[ 1-\exp \left( -\frac \omega {T_{{\rm BH}}}\right) \right] \omega  \over \left[ 1-\exp \left( -\frac \omega {2T_R}\right) \right] \left[ 1-\exp \left( -\frac \omega {2T_L}\right) \right] }}
, & 
{\displaystyle {1 \over T_{{\rm BH}}}}
=%
{\displaystyle {1 \over 2}}
\left( 
{\displaystyle {1 \over T_R}}
+%
{\displaystyle {1 \over T_L}}
\right)
\end{array}
\label{aru.118.p}
\end{equation}
(this relationship agrees with (4.34) of \cite{Malda.grey}).

The result found for the D-string can now be easily generalized to the four
and five dimensional black holes studied in \cite{Malda} and \cite
{Malda.grey}: the change is simply in the central charge of the SCFT. In the
case of the five dimensional black hole the central charges are ${\sf C}_{%
{\rm free}}^{{\rm D}\left( 1+5\right) }={\sf \tilde C}_{{\rm free}}^{{\rm D}%
\left( 1+5\right) }=6Q_1Q_5$, while for the four dimensional black hole, one
has: ${\sf C}_{{\rm free}}^{{\rm D}\left( 2+6\right) {\rm -}5{\rm s}}={\sf 
\tilde C}_{{\rm free}}^{{\rm D}\left( 2+6\right) {\rm -}5{\rm s}}=6Q_2Q_5Q_6$
\cite{malda.priv.1}.

\section{The Hawking temperature law.\label{QUATESE-HAWKING}}

The emergence of a Hawking effect is explained in terms of the huge
degeneracy of the quantum levels for the ${\frak Vir}$- or ${\frak Vir}%
\oplus \overline{{\frak Vir}}$-symmetric systems, and it takes place when
the conditions (\ref{limite.canonico}) are met. So there must be the way to
predict the value of this temperature simply by assigning the conformal
symmetry basics of the system at hand: here I suggest a law expressing the
Hawking temperature as a function of the central charges only.

In the case of a ${\frak Vir}$-symmetric theory, if ${\sf C}_{{\rm free}}$
is the value of the central charge, there is a QUATESE temperature, reading: 
\begin{equation}
T=K\sqrt{\frac 6{{\sf C}_{{\rm free}}}};  \label{temps.1}
\end{equation}
here $K$ is a {\it constant}.

In the case of the ${\frak Vir}\oplus \overline{{\frak Vir}}$-symmetric
systems there exist two QUATESE temperatures, on left and one right-moving,
which are still expressed as: 
\begin{equation}
\begin{array}{cc}
T_R=K\left( N\right) \sqrt{%
{\displaystyle {6 \over {\sf C}_{{\rm free}}}}
}, & T_L=K\left( N\right) \sqrt{%
{\displaystyle {6 \over {\sf \tilde C}_{{\rm free}}}}
}.
\end{array}
\label{temps.1.bis}
\end{equation}
The spectrum is a grey body one; here $K$ is a {\it function of the mass
quantum level}, around which the emitting transition takes place.

In what follows the expressions corresponding to this equation (\ref{temps.1}%
) and (\ref{temps.1.bis}) are singled out for systems showing quasi-thermal
emission: the F- and D-strings and several kinds of black hole, and the
functions $K\left( N\right) $ are put in evidence.

\subsection{Strings and string-theoretical black holes.}

The first system is the {\it F-string}. In the open case, we can single out
the function $K$ as: 
\begin{equation}
\begin{array}{cccc}
T_{{\rm open}}^{\left( {\rm F}\right) }=%
{\displaystyle {1 \over 2\pi \ell _s}}
\sqrt{%
{\displaystyle {6 \over {\sf C}_{{\rm free}}}}
} & \Rightarrow & K_{{\rm open}}^{\left( {\rm F}\right) }=%
{\displaystyle {1 \over 2\pi \ell _s}}
, & 
{\displaystyle {\partial \over \partial N}}
K_{{\rm open}}^{\left( {\rm F}\right) }=0.
\end{array}
\label{temps.2.ter}
\end{equation}

In the case of the {\it closed} F-string, there is {\it a right and a left
moving temperatures}: the functions $K$ read: 
\begin{equation}
\begin{array}{cc}
K_R^{\left( {\rm F}\right) }=%
{\displaystyle {\sqrt{N_{R,L}-1} \over \pi \ell _s^2M}}
, & K_L^{\left( {\rm F}\right) }=%
{\displaystyle {\sqrt{N_{R,L}-1} \over \pi \ell _s^2M}}
.
\end{array}
\label{temps.4}
\end{equation}
In the {\it long string approximation} $T^{\left( {\rm F}\right) }L\gg 1$
these read: 
\begin{equation}
K_{R,L}^{\left( {\rm F}\right) }\left( N\right) =%
{\displaystyle {1 \over \pi R}}
\sqrt{N_{R,L}-1}.  \label{temps.6.bis}
\end{equation}

When the inclusive calculation is performed for the {\it D-string}, the
functions $K\left( N\right) $ are equal as in the case of the F-string (\ref
{temps.6.bis}): 
\begin{equation}
K_{R,L}^{\left( D\right) }\left( N\right) =%
{\displaystyle {1 \over \pi R}}
\sqrt{N_{R,L}-1}.  \label{temps.7.bis}
\end{equation}

The {\it five dimensional near extreme black} hole is represented by a D$%
\left( 1+5\right) $-brane bound system \cite{Malda}: when it is treated with 
{\it canonical ensemble calculations}, the values of right and left
temperatures read 
\begin{equation}
\begin{array}{cc}
T_R^{{\rm D}\left( 1+5\right) }\left( {\rm c.e.c.}\right) =%
{\displaystyle {1 \over \pi R}}
\sqrt{%
{\displaystyle {6N_R \over {\sf C}^{{\rm D}\left( 1+5\right) }}}
}, & T_L^{{\rm D}\left( 1+5\right) }\left( {\rm c.e.c}\right) =%
{\displaystyle {1 \over \pi R}}
\sqrt{%
{\displaystyle {6N_L \over {\sf \tilde C}^{{\rm D}\left( 1+5\right) }}}
}.
\end{array}
\label{temps.12}
\end{equation}
By evaluating the {\it inclusive emission rate} of the D$\left( 1+5\right) $%
-system, a quasi thermal spectrum is found, with the QUATESE temperatures 
\begin{equation}
\begin{array}{cc}
T_R^{{\rm D}\left( 1+5\right) }\left( {\rm i.e.r.}\right) =%
{\displaystyle {1 \over \pi R}}
\sqrt{%
{\displaystyle {6N_R \over {\sf C}^{{\rm D}\left( 1+5\right) }}}
}, & T_L^{{\rm D}\left( 1+5\right) }\left( {\rm i.e.r.}\right) =%
{\displaystyle {1 \over \pi R}}
\sqrt{%
{\displaystyle {6N_L \over {\sf \tilde C}^{{\rm D}\left( 1+5\right) }}}
}.
\end{array}
\label{temps.13}
\end{equation}

If one deals with the {\rm D}$\left( 2+6\right) $-$5s$-system describing the
SUGRA four dimensional black hole in \cite{Malda}, the QUATESE temperatures
would follow again the law (\ref{temps.1}) in the form 
\begin{equation}
\begin{array}{cc}
T_R^{{\rm D}\left( 2+6\right) -5s}\left( {\rm i.e.r.}\right) =%
{\displaystyle {1 \over \pi R}}
\sqrt{%
{\displaystyle {6N_R \over {\sf C}^{{\rm D}\left( 2+6\right) -5s}}}
}, & T_L^{{\rm D}\left( 2+6\right) -5s}\left( {\rm i.e.r.}\right) =%
{\displaystyle {1 \over \pi R}}
\sqrt{%
{\displaystyle {6N_L \over {\sf \tilde C}^{{\rm D}\left( 2+6\right) -5s}}}
},
\end{array}
\label{temps.13.bis}
\end{equation}
being ${\sf C}^{{\rm D}\left( 2+6\right) -5s}={\sf \tilde C}^{{\rm D}\left(
2+6\right) -5s}=6Q_2Q_5Q_6$, and the canonical ensemble $T_{R,L}^{{\rm D}%
\left( 2+6\right) -5s}\left( {\rm c.e.}\right) $ would have the same form.

\subsection{General relativity black holes.\label{Carlip}}

Someone has found a ${\frak Vir}$-symmetry which characterizes the physics
of a classical Einstein black hole in any dimension, arbitrarily far from
extremality: this has been done by Carlip (see \cite{Carlip.3} and \cite
{Carlip.4}), who showed that {\it the boundary physics of a stationary
spacetime with an horizon is invariant under a class of diffeomorphisms of
the Cauchy hypersurfaces, forming a Virasoro algebra with central charge} 
\begin{equation}
{\sf C}_{{\rm BH}}=\frac{3{\cal A}}{2\pi G},  \label{temps.15}
\end{equation}
being ${\cal A}$ the measure of the bifurcation sphere at the horizon.

Does the relationship (\ref{temps.1}) apply to those black holes considered
by Carlip, when the role of the central charge is played by $\frac{3{\cal A}%
}{2\pi G}$, as equation (\ref{temps.15})\ predicts?

Let us consider the {\it Schwarzschild classical black hole}. The horizon
area reads 
\begin{equation}
\begin{array}{ccc}
{\cal A}=16\pi G^2M^2 & \Rightarrow & {\sf C}_{{\rm BH}}=24GM^2,
\end{array}
\label{temps.16}
\end{equation}
while the Hawking temperature is 
\begin{equation}
T_{{\rm BH}}^{{\rm Schw}}=\frac 1{8\pi GM}:  \label{temps.17}
\end{equation}
then one has the following relationship 
\begin{equation}
\begin{array}{cc}
T_{{\rm BH}}^{{\rm Schw}}=%
{\displaystyle {1 \over 4\pi \sqrt{G}}}
\sqrt{%
{\displaystyle {6 \over {\sf C}_{{\rm BH}}}}
}, & K_{{\rm BH}}^{{\rm Schw}}=%
{\displaystyle {1 \over 4\pi \sqrt{G}}}
\end{array}
\label{temps.18}
\end{equation}
thus, the Hawking temperature of a Schwarzschild black hole does obey the
law (\ref{temps.1}).

Consider the {\it Reissner-Nordstr\"om classical black hole}:\ from (\ref
{temps.15}) one finds 
\begin{equation}
{\sf C}_{{\rm BH}}=%
{\displaystyle {6r_{+}^2 \over G}}
=6G\left( 2M^2-Q^2+2M\sqrt{M^2-Q^2}\right) .  \label{temps.19}
\end{equation}
The Hawking temperature for the Reissner-Nordstr\"om solution reads: 
\begin{equation}
T_{{\rm BH}}^{{\rm R-N}}= 
{\displaystyle {\sqrt{G^2M^2-Q^2} \over 2\pi \sqrt{G}\left( GM+\sqrt{G^2M^2-Q^2}\right) }}
\sqrt{%
{\displaystyle {6 \over {\sf C}_{{\rm BH}}}}
}.  \label{temps.21}
\end{equation}

To trace a relationship between the emitting strings and the charged black
hole of General relativity, a comparison between slightly non BPS systems
has to be made, hoping that this will quantum-protect our calculations. For 
{\it small values} of $\sqrt{G^2M^2-Q^2}$, the factor $K_{{\rm R-N}}$ is
expanded as: 
\begin{equation}
K_{{\rm R-N}}\left( M\right) = 
{\displaystyle {\sqrt{G^2M^2-Q^2} \over 2\pi \sqrt{G}\left( GM+\sqrt{G^2M^2-Q^2}\right) }}
=%
{\displaystyle {\sqrt{G^2M^2-Q^2} \over 2\pi G^{\frac 32}M}}
+...  \label{temps.21.1}
\end{equation}
If some law as (\ref{aru.72}) is assumed 
\[
\begin{array}{cc}
G^2M^2-Q^2=\alpha _{{\rm R-N}}^2N, & N\in {\Bbb N},
\end{array}
\]
in the spirit of \cite{susskind.1}, then the relationship (\ref{temps.21.1})
becomes: 
\begin{equation}
K_{{\rm R-N}}\left( M\right) =%
{\displaystyle {\alpha _{{\rm R-N}} \over 2\pi G^{\frac 32}M}}
\sqrt{N}+...,  \label{temps.21.2}
\end{equation}
which is quite similar to (\ref{temps.7.bis}) (this $\alpha _{{\rm R-N}}$ is
some constant introduced {\it ad hoc}). This relationship (\ref{temps.21.2})
suggests that {\it nearly extremal Reissner-Nordstr\"om black holes show a
Hawking temperature that obeys the QUATESE law (\ref{temps.1.bis}) if
Carlip's interpretation of the area law is adopted.}

\section{Conclusions.}

The emission rate from F-strings becomes approximately thermal when very
excited states are treated, and low energy particles emitted: if the rate is
summed over the final states and averaged over the initial ones, then a
Planckian distribution appears.

Can string theoretical black holes benefit of a similar mechanism? The
answer is yes, because the string modeled black holes studied here can be
represented as a suitable $1+1$ SCFT: this is explicitly verified for
D-strings, for five dimensional holes corresponding to D$\left( 1+5\right) $
systems, and for four dimensional holes corresponding to D$\left( 2+6\right) 
$-$5$s systems. In all these cases, the emission temperature is a function
of the central charge of the SCFT at hand, which has the same forms (\ref
{temps.1}) and (\ref{temps.1.bis}) for all these systems.

The emission rates have to be inclusive and averaged, when black holes are
treated, due to the presence of the event horizon that prevents external
observers from seeing the exact state of the black hole, even if its ADM
charges can be measured from infinity.

The QUATESE mechanism seems to give the right answer for the Hawking
temperature even outside the full superstring theory, as it is shown when
the Schwarzschild and the near extremal Reissner-Nordstr\"om black holes of
General relativity are examined in the spirit of \cite{Carlip.4}.

The natural development of this result is to check what happens when a
string falls into a black hole whose horizon physics is described by
Carlip's Virasoro algebra: is the central charge of the worldsheet ${\frak %
Vir}\oplus \overline{{\frak Vir}}$ algebra of the F-string related to the
increment $\delta {\cal A}$ of the measure of the bifurcation sphere, due to
its absorption? Is the worldsheet ${\frak Vir}\oplus \overline{{\frak Vir}}$
algebra of every fundamental string physically ''the same'' algebra
characterizing the black hole?

In case of a positive answer, this line of research would allow to claim for
the stringy nature of elementary constituents of matter simply due to the
existence of black hole, because only one dimensional objects have Virasoro
algebr\ae\ among their internal symmetries, and in this vision the $1+1$
conformal symmetry would be promoted to the role of spacetime fundamental
symmetry.

\bigskip\ 

\begin{center}
{\bf Acknowledgments}
\end{center}

I would like to thank Professor Giorgio Immirzi of Perugia University, whom
I have been working with during my PhD period, and Professor Mario Calvetti
of Florence University, who encouraged me to publish these results.

A special thank goes to Silvia Aldrovandi, who never stopped trusting in me.

\end{document}